\documentclass[sigconf,nonacm]{acmart}  %
\usepackage{xspace}
\usepackage[noabbrev,capitalise]{cleveref}
\usepackage{siunitx}
\usepackage{subcaption}

\newcommand\blfootnote[1]{
    \begingroup
    \renewcommand\thefootnote{}\footnote{#1}
    \addtocounter{footnote}{-1}
    \endgroup
}

\usepackage[nolist]{acronym} %

\AtBeginDocument{%
  }

\copyrightyear{2024}
\acmYear{2024}
\setcopyright{rightsretained}
\acmConference[RAID 2024]{The 27th International Symposium on Research in Attacks, Intrusions and Defenses}{September 30-October 02, 2024}{Padua, Italy}
\acmBooktitle{The 27th International Symposium on Research in Attacks, Intrusions and Defenses (RAID 2024), September 30-October 02, 2024, Padua, Italy}
\acmDOI{10.1145/3678890.3678922}
\acmISBN{979-8-4007-0959-3/24/09}

\renewcommand{\arraystretch}{1.35}

\newcommand{\datasetLabelFake}{\ensuremath{\mathcal{D}_F}\xspace}
\newcommand{\datasetLabelFakeX}{\ensuremath{\mathcal{D}_F^\mathbb{X}}\xspace}
\newcommand{\datasetLabelFakeZoomed}{\ensuremath{\mathcal{D}_F^{\mathbb{X}'}}\xspace}
\newcommand{\datasetLabelReal}{\ensuremath{\mathcal{D}_R}\xspace}
\newcommand{\datasetLabelRealX}{\ensuremath{\mathcal{D}_R^\mathbb{X}}\xspace}
\newcommand{\datasetLabelRealZoomed}{\ensuremath{\mathcal{D}_R^{\mathbb{X}'}}\xspace}
\newcommand{\datasetInWild}{\ensuremath{\mathcal{D}_W^{\mathbb{X}}}\xspace}
\newcommand{\datasetPseudoLabel}{\ensuremath{\mathcal{D}_P^\mathbb{X}}\xspace}
\newcommand{\datasetSpotted}{\ensuremath{\mathcal{D}_D^\mathbb{X}}\xspace}

\newcommand{\prefilter}{\ensuremath{\phi}\xspace}

\newcommand{\classifier}{\ensuremath{\mathcal{C}}\xspace}
\newcommand{\classifierRF}{\ensuremath{\mathcal{C}_{R/F}}\xspace}
\newcommand{\classifierRxFx}{\ensuremath{\mathcal{C}_{R^\mathbb{X} / F^\mathbb{X}}}\xspace}
\newcommand{\classifierRxPxFx}{\ensuremath{\mathcal{C}_{R^\mathbb{X}, P^\mathbb{X} / F^\mathbb{X}}}\xspace}
\newcommand{\classifierWang}{\ensuremath{\mathcal{C}_\text{Wang}}\xspace}
\newcommand{\classifierGrag}{\ensuremath{\mathcal{C}_\text{Grag}}\xspace}
\newcommand{\classifierOjha}{\ensuremath{\mathcal{C}_\text{Ojha}}\xspace}

\renewcommand{\paragraph}[1]{{\vskip 6pt \noindent\textbf{#1.} }}

\newcommand{\eg}{e.g.,\xspace} %
\newcommand{\cf}{cf.\xspace} %

\begin{acronym}
    \acro{AIGC}{AI-generated content}
    \acro{FNR}{false negative rate}
    \acro{FDR}{false discovery rate}
    \acro{TPR}{true positive rate}
    \acro{FPR}{false positive rate}
    \acro{GAN}{generative adversarial network}
    \acro{TPDNE}{\href{https://thispersondoesnotexist.com}{thispersondoesnotexist.com}}
    \acro{TF-IDF}{term frequency–inverse document frequency}
\end{acronym}

\begin{document}

\title[AI-Generated Faces in the Real World: A Large-Scale Case Study of Twitter Profile Images]{AI-Generated Faces in the Real World: \\A Large-Scale Case Study of Twitter Profile Images}

\author{Jonas Ricker}
\orcid{0000-0002-7186-3634}
\affiliation{%
  \institution{Ruhr University Bochum}
  \city{Bochum}
  \country{Germany}
}

\author{Dennis Assenmacher}
\orcid{0000-0001-9219-1956}
\affiliation{%
 \institution{GESIS - Leibniz Institute for the Social Sciences}
 \city{Cologne}
 \country{Germany}
}

\author{Thorsten Holz}
\orcid{0000-0002-2783-1264}
\affiliation{%
 \institution{CISPA Helmholtz Center for Information Security}
 \city{Saarbrücken}
 \country{Germany}
}
\authornote{Equal supervision.}

\author{Asja Fischer}
\orcid{0000-0002-1916-7033}
\affiliation{%
\institution{Ruhr University Bochum}
\city{Bochum}
\country{Germany}
}
\authornotemark[1]

\author{Erwin Quiring}
\orcid{0009-0004-7170-1274}
\affiliation{%
\institution{Ruhr University Bochum}
\city{Bochum}
\country{Germany}
}
\affiliation{%
\institution{ICSI}
\city{Berkeley}
\country{USA}
}
\authornotemark[1]

\begin{abstract}
Recent advances in the field of generative artificial intelligence (AI) have blurred the lines between authentic and machine-generated content, making it almost impossible for humans to distinguish between such media.
One notable consequence is the use of AI-generated images for fake profiles on social media. 
While several types of disinformation campaigns and similar incidents have been reported in the past, a systematic analysis has been lacking. 

In this work, we conduct the first large-scale investigation of the prevalence of AI-generated profile pictures on Twitter.
We tackle the challenges of a real-world measurement study by carefully integrating various data sources and designing a multi-stage detection pipeline. 
Our analysis of nearly 15 million Twitter profile pictures shows that \SI{0.052}{\percent} were artificially generated, confirming their notable presence on the platform. 
We comprehensively examine the characteristics of these accounts and their tweet content, and uncover patterns of coordinated inauthentic behavior. 
The results also reveal several motives, including spamming and political amplification campaigns. 
Our research reaffirms the need for effective detection and mitigation strategies to cope with the potential negative effects of generative AI in the future. %
\blfootnote{© 2024 Copyright held by the owner/author(s). This is the author's version of the work. It is posted here for your personal use. Not for redistribution. The definitive Version of Record was published in RAID '24: Proceedings of the 27th International Symposium on Research in Attacks, Intrusions and Defenses, \url{https://dx.doi.org/10.1145/3678890.3678922}.}
\end{abstract}

\ccsdesc[500]{Computing methodologies~Machine learning}
\ccsdesc[500]{Information systems~Social networks}
\ccsdesc[500]{Security and privacy~Human and societal aspects of security and privacy}

\keywords{AI-Generated Content, Fake Image Detection, Social Networks}

\maketitle

\section{Introduction}\label{sec:introduction}

The emergence of generative artificial intelligence (AI) has revolutionized content creation, enabling us to produce highly authentic and diverse outputs, such as images, videos, texts, and music that bear a striking resemblance to human-created media. 
These AI-driven systems have become ubiquitous in various areas of our society and provide reliable support in numerous applications.
Among other use cases, they streamline the writing of emails and texts or enhance programming with advanced code completion tools. 
However, alongside its impressive benefits, generative AI also has the potential for significant detrimental effects. 
A pressing problem %
is the ability to generate compellingly realistic but false content, which can be used as a way to spread misinformation, manipulate people, and influence public opinion.

In a significant action in late 2019, Facebook dismantled an extensive network of over 900 accounts, pages, and groups that had collectively spent more than 9 million USD on advertisements promoting Donald Trump, potentially impacting the 2020 US presidential election~\citep{nimmoOperationFFSFakeFace2019}. 
A notable feature of this network was the use of AI-generated profile images, possibly taken from the website \ac{TPDNE} which became operational in February 2019. 
Using NVIDIA's StyleGAN~\citep{karrasStylebasedGeneratorArchitecture2019}, TPDNE generates a new facial image every time the page is refreshed, making it easily accessible to everyone.
Since this incident, the use of AI-synthesized faces in disinformation campaigns has been on the rise, likely because such images reduce the risk of detection through reverse image searches~\citep{goldsteinHowDisinformationEvolved2021}. 
Investigations have revealed that many of these deceptive clusters operate with state interests in mind, seeking to bolster specific narratives~\citep{nimmoOperationNavalGazing2020,nimmoSpamouflageGoesAmerica,strickAnalysisProchinaPropaganda2021,stanfordinternetobservatoryAnalysisTwitterTakedowns2020} or interfere in the domestic policies of foreign states~\citep{nimmoIRAAgainUnlucky2020,graphikateamStepMyParler2020,graphikateamFakeClusterBoosts2021}. 
Additionally, there are efforts to influence public opinion~\citep{stanfordinternetobservatoryReplyguysGoHunting2020,strickWestPapuaNew2020} or establish connections with unsuspecting social media users~\citep{vincentSpyReportedlyUsed2019,goldsteinResearchNoteThis2022}. 
The FBI and Europol have expressed concerns that the trend of using AI-generated content in cybercrime and foreign influence operations is expected to grow steadily~\citep{MaliciousActorsAlmost2021,Europol2024}.
Given these examples, it is essential to understand the detection possibility, prevalence, and usage of AI-generated images in the wild instead of a lab setting. 

In this work, we tackle this challenge by concentrating on the phenomenon of AI-generated images in social media. At the time of writing, it is becoming increasingly difficult for humans to differentiate these machine-generated media from authentic photographs, as evidenced by recent studies~\citep{hulzeboschDetectingCNNgeneratedFacial2020, tucciarelliRealnessPeopleWho2020, nightingaleSyntheticFacesHow2021, shenStudyHumanPerception2021, lagoMoreRealReal2022, nightingaleAIsynthesizedFacesAre2022, frankRepresentativeStudyHuman2023}.
Although the detection of generated images has been explored extensively in lab settings, there is a surprising lack of comprehensive research addressing their identification and widespread use on social media platforms in real-world contexts.
In this paper, we provide the first systematic and large-scale study of AI-generated profile images on Twitter\footnote{Since we collected our main dataset before Twitter's rebranding to $\mathbb{X}$, we use the platform's former name throughout this work.}. 
Our research is founded on three main pillars.
 
First, we develop a fast and effective detection pipeline tailored to the identification of AI-generated images in real-world scenarios. 
This task presents unique challenges, including the lack of a definitive ground truth and the diversity of possible image manipulations.
To solve these problems, we carefully design a detection pipeline step by step. %
We consider different dataset types, apply a pre-filter to discard images with too small or no faces, and adapt a state-of-the-art classification model specifically targeting synthetic profile images on Twitter. %
As mentioned above, observations suggest that the majority of AI-generated profile images originate from TPDNE, which is why we tailor our detection pipeline to this kind of fake faces.
Finally, we integrate various tools that help with the manual labeling that is required to estimate error rates on unlabeled in-the-wild data.
We study each component of our system in controlled setups and show that the pipeline is capable of accurately recognizing AI-generated images. 

Second, we analyze a large collection of \num{14989385} Twitter profile pictures to determine how prevalent AI-generated profile pictures are on the platform. 
We identify \num{7723} accounts that use such images, which corresponds to a prevalence rate of \SI{0.052}{\percent}. 
This result indicates a notable presence of generated profile images on Twitter.
We also assess the accuracy and reliability of our findings by estimating error rates. %
We estimate the \ac{FNR}---the fraction of mislabeled fake images---of our approach to lie between \SI{2.75}{\percent} and \SI{3.03}{\percent}, and the \ac{FDR}---the fraction of real images among all images classified as fake---to be \SI{1.4}{\percent}. 
The results suggest a low error rate of our method. %

Third, we contextualize the use of AI-generated profile pictures on Twitter by examining the corresponding accounts and their tweets. 
Our results show clear differences between the two types of accounts: accounts with fake images tend to have lower social engagement as well as fewer followers and followed accounts. 
Despite the generally lower activity, some accounts with fake images are very active, suggesting possible involvement in spam campaigns. 
In addition, fake accounts are often newer and are suspended more frequently by Twitter, indicating inauthentic behavior. 
A significant portion of accounts was created in bulk shortly before our data collection, which is a common pattern for accounts created for message amplification, disinformation campaigns, or similar disruptive activity.
This impression is confirmed by our textual analysis of the accounts' tweets. We identify large clusters spamming very similar contents, frequently referring to giveaways, cryptocurrencies, and pornography. Notably, we also observe accounts that engage in contentious or political topics, such as the war in Ukraine, debates on COVID-19 and vaccinations, and election-related discourse.

\paragraph{Contributions}
We make the following key contributions:
\begin{enumerate}
    \setlength{\itemsep}{3pt}
    \item {\em Detection Pipeline.}\quad We propose a multi-step pipeline for detecting AI-generated profile images on social media. We evaluate each stage in a controlled setup and demonstrate the pipeline's suitability for real-world settings.
    \item \emph{Prevalence Study on Twitter.}\quad We apply our pipeline to a total of \num{14989385} authentic profile images to systematically study the prevalence of AI-generated faces on Twitter. We identify \num{7723} accounts with generated profile images, corresponding to a prevalence rate of \SI{0.052}{\percent}.
    \item \emph{Account and Tweet Analysis.}\quad We analyze the user metrics and tweets of accounts using AI-generated profile images to learn more about their intended purpose. We identify prevalent topics and find a significant number of accounts to apparently participate in coordinated inauthentic behavior.
\end{enumerate}
We release our code and data at \href{https://github.com/jonasricker/twitter-ai-faces}{github.com/jonasricker/twitter-ai-faces}.

\section{Background}\label{sec:background}
We start by providing a short primer on the creation and detection of AI-generated images. %

\paragraph{AI-Generated Content (AIGC)}\label{sec:background:deepfakes}
\acs{AIGC}, sometimes also referred to as ``deepfakes'', is content that appears authentic to humans but is synthesized or altered using a deep neural network.
It is most prominently associated with manipulated videos in which the face of a person is replaced with a different one~\citep{mirskyCreationDetectionDeepfakes2021}, but also encompasses other types of media including images, audio, and text.
While AIGC offers great creative potential, it is also used for malicious purposes, including defamatory images and videos~\citep{coleAIassistedFakePorn2017}, voice cloning~\citep{gaoVoiceImpersonationUsing2018,damianiVoiceDeepfakeWas}, fake customer reviews~\citep{yaoAutomatedCrowdturfingAttacks2017}, and machine-generated posts on social media~\citep{fagniTweepFakeDetectingDeepfake2021,goldsteinGenerativeLanguageModels2023}.

\paragraph{Image Synthesis}\label{sec:background:deepfake_generation}
Learning a probability distribution from samples in order to generate novel samples is a longstanding challenge, 
especially in the high-dimensional image domain.
Besides variational autoencoders (VAEs)~\citep{kingmaAutoencodingVariationalBayes2014} and autoregressive models~\citep{oordPixelRecurrentNeural2016a,vandenoordConditionalImageGeneration2016}, \acp{GAN}~\citep{goodfellowGenerativeAdversarialNets2014} have proven to be effective in synthesizing high-quality images~\citep{zhuUnpairedImagetoimageTranslation2017,choiStarGANUnifiedGenerative2018,karrasProgressiveGrowingGANs2018,karrasStylebasedGeneratorArchitecture2019,karrasAnalyzingImprovingImage2020,karrasAliasFreeGenerativeAdversarial2021,sauerProjectedGANsConverge2021,kangScalingGANsTexttoimage2023}.
The StyleGAN family~\citep{karrasStylebasedGeneratorArchitecture2019,karrasAnalyzingImprovingImage2020,karrasAliasFreeGenerativeAdversarial2021} received special attention due to their ability to generate faces that are practically indistinguishable from real ones~\citep{nightingaleAIsynthesizedFacesAre2022}.
Recently, it has been shown that diffusion models (DMs)~\citep{sohl-dicksteinDeepUnsupervisedLearning2015,hoDenoisingDiffusionProbabilistic2020,dhariwalDiffusionModelsBeat2021} are able to match and even surpass the visual quality of GAN-generated images.

\paragraph{Generated Image Detection}\label{sec:background:deepfake_detection}
There is a continuing arms race for effective detection techniques and newer generations of image synthesis algorithms.
Broadly speaking, generated image detection techniques can be divided into two categories: methods that rely on handcrafted features and learning-based methods.
Methods from the first category either exploit visual defects (\eg facial inconsistencies~\citep{maternExploitingVisualArtifacts2019}, impossible reflections~\citep{huExposingGANGeneratedFaces2021}, irregular pupil shapes~\citep{guoEyesTellAll2022}) or ``invisible'' characteristics such as frequency artifacts~\citep{zhangDetectingSimulatingArtifacts2019,durallWatchYourUpconvolution2020,frankLeveragingFrequencyAnalysis2020,chandrasegaranCloserLookFourier2021,schwarzFrequencyBiasGenerative2021,chenSSDGANMeasuringRealness2021}, pixel statistics~\citep{natarajDetectingGANGenerated2019,mccloskeyDetectingGANgeneratedImagery2019}, or model-specific properties~\citep{marraGANsLeaveArtificial2019,yuAttributingFakeImages2019,rickerAEROBLADE2024}.
Learning-based methods, on the other hand, use neural networks to learn a suitable feature representation to distinguish fake from real images~\citep{marraDetectionGANgeneratedFake2018,chaiWhatMakesFake2020,hulzeboschDetectingCNNgeneratedFacial2020,wangCNNgeneratedImagesAre2020,gragnanielloAreGANGenerated2021,cozzolinoUniversalGANImage2021,ojhaUniversalFakeImage2023,corviDetectionSyntheticImages2023}.

\section{Methodology}\label{sec:method}

A large-scale study on generated images in the wild comes with multiple challenges. First, we do not know the ground truth. As a result, it is difficult to estimate the amount of overlooked generated images (\emph{false~negatives}) and to be sure that an image detected as generated is actually generated (\emph{precision}). 
Finally, studying millions of images comes with a computational overhead so that the detection method has to be efficient, too. We discuss further challenges and limitations of our study in Section~\ref{sec:discussion}. 
To deal with all these challenges, we carefully design a multi-step detection pipeline. The following is a step-by-step description of this pipeline. Note that while the presented approach is applied to Twitter, our method can be adapted to any other social network. We provide implementation details in Appendix~\ref{app:implementation}.

\subsection{Data Collection}\label{sec:method:data_collection}
We describe the four types of datasets that we use for studying generated images, with Twitter being our use case. Table~\ref{tab:symbols} summarizes our notation.

\paragraph{In-The-Wild Dataset \datasetInWild} 
To estimate the prevalence of generated images on a social network, it is important to obtain a mostly unconditional sample.
In the case of Twitter, this can be achieved by using the API endpoint that provides real-time access to a random \SI{1}{\percent}~subset of all publicly posted tweets.
We download each author's profile image together with their profile metadata (\cf\ Appendix~\ref{app:metadata} for an overview).
Note that this approach only enables us to obtain profile images from users who write posts during the data collection period.
Additionally, we omit users who have not set a profile image, that is, who are using Twitter's default profile image.
From March~7 to March~15 2023, we collected \num{14989385} profile images.

\paragraph{Labeled Datasets \datasetLabelReal/\datasetLabelFake and Variations}
We continue with labeled datasets of fake and real images which can be used to train a detector. %
As discussed in Section~\ref{sec:introduction}, existing observations suggest that the vast majority of generated profile images on Twitter are taken from 
TPDNE, which generates images with StyleGAN2~\citep{karrasAnalyzingImprovingImage2020} trained on the FFHQ~\citep{karrasStylebasedGeneratorArchitecture2019} dataset\footnote{When published in 2019, TPDNE used the original StyleGAN~\citep{karrasStylebasedGeneratorArchitecture2019}, but switched to StyleGAN2~\citep{karrasAnalyzingImprovingImage2020} shortly after its release.}.
We therefore decide to focus on this specific kind of fake faces and use \num{10000} images from TPDNE as our fake-labeled dataset (denoted by~\datasetLabelFake) and correspondingly \num{10000} images from FFHQ as our real-labeled dataset (denoted by~\datasetLabelReal). 
We discuss this limitation of focusing on TPDNE in Section~\ref{sec:discussion}.
As prior work shows that processing operations like resizing and compression can affect the detection~\citep{parmarOnAliasedResizing2022, mandelliTrainingCNNsPresence2020}, we consider two dataset variations:
\begin{itemize}
    \setlength\itemsep{0em}
    \item \datasetLabelRealX and \datasetLabelFakeX. To obtain profile images with the social network's processing steps, we adapt the approach from \citet{boatoTrueFaceDatasetDetection2022}. We upload both \datasetLabelReal and \datasetLabelFake to Twitter, set each image as profile image, and then download all images again. We denote these processed images by~\datasetLabelRealX and \datasetLabelFakeX.
    \item \datasetLabelRealZoomed and \datasetLabelFakeZoomed. We additionally simulate a user who zooms into the profile image during the upload, as it is common for social media platforms. We denote these images by~\datasetLabelRealZoomed and \datasetLabelFakeZoomed, respectively.
\end{itemize}
We confirm in Section~\ref{sec:detector_ablation} that considering the preprocessing indeed improves the detection performance under realistic conditions.

\paragraph{Proxy-Labeled Real Dataset~\datasetPseudoLabel}
Social media platforms often have very popular users with a lot of followers. These popular users are rather unlikely to use deceptive fake images. Hence, we can build a proxy-labeled dataset with presumably \emph{real} images. In particular, we select \num{10000} profile images from the accounts in \datasetInWild with the highest numbers of followers that also pass our pre-filter (which is presented in the next section).
We denote the so-created proxy-labeled dataset of real profile images by~\datasetPseudoLabel. 

\paragraph{Documented Fakes Dataset~\datasetSpotted}
Finally, there are documented cases of generated profile images that were discovered manually. For example, blog posts regularly report such images when analyzing inauthentic Twitter accounts~\citep{nortenoConspiradorNortenoSubstack2024}. These cases can be used to build a labeled dataset of fake images in the wild, which we denote by~\datasetSpotted. Such a dataset is not free of bias, but provides a good means to finally check the performance of our classifier on an independent source.
For our study, we use a dataset of \num{1420} generated Twitter profile images that were manually collected between November 2022 and January 2024~\citep{yangCharacteristicsPrevalenceFake2024}.

\begin{table}[t!]
\centering
\footnotesize
\caption{Dataset notation. The symbol $^\mathbb{X}$ indicates that images were processed by Twitter.}
\label{tab:symbols}
\begin{tabular}{|l||p{6.5cm}|}
  \hline
  {\sc Symbol} & {\sc Description} \\
  \hline \hline
  $\datasetInWild$ & Unlabeled dataset of Twitter profile images. \\ \hline
  $\datasetLabelFake$ & Labeled dataset of fake images. \\ \hline
  $\datasetLabelFakeX$ & Labeled dataset of fake images uploaded as profile image and downloaded afterward. \\ \hline
  $\datasetLabelFakeZoomed$ & Version of $\datasetLabelFakeX$ where images are zoomed into during upload. \\ \hline  
  $\datasetLabelReal$ & Labeled dataset of real images. \\ \hline
  $\datasetLabelRealX$ & Labeled dataset of real images uploaded as profile image and downloaded afterward. \\ \hline
  $\datasetLabelRealZoomed$ & Version of $\datasetLabelRealX$ where images are zoomed into during upload. \\ \hline
  $\datasetPseudoLabel$ & Proxy-labeled dataset of supposedly real Twitter profile images. \\ \hline
  $\datasetSpotted$ & Labeled dataset of documented fake Twitter profile images. \\ \hline
\end{tabular}
\end{table}

\subsection{Detection}\label{sec:method:detection}
Equipped with these different datasets, we can proceed with the detection of generated profile images. Here, we propose a two-stage procedure to improve the accuracy and the efficiency.  

\paragraph{Pre-Filter \prefilter}
We start with a pre-filter~\prefilter to discard irrelevant samples.
In our case, we can discard images without any face or where the face is too small.
We use the efficient BlazeFace~\citep{bazarevskyBlazeFace2019} face detector to detect faces and locate facial landmarks.
An image passes \prefilter if at least one face is detected and the Euclidean distance between the normalized coordinates of both eyes is greater or equal to~\num{0.1}.
The pre-filter serves two purposes: First, the overall computational complexity decreases by reducing the number of analyzed candidates in the subsequent, more demanding detection stage.
Second, the detection stage is trained on facial images, so that other types of profile images, such as logos or monochrome images, could be wrongly classified as fake.
Filtering irrelevant images can therefore decrease the \ac{FPR}.

\paragraph{Classifier \classifier}
To automatically label a profile image as real or fake, we use a state-of-the-art CNN detector based on ResNet-50~\citep{heDeepResidualLearning2016}. Previous work~\citep{wangCNNgeneratedImagesAre2020, mandelliTrainingCNNsPresence2020, cozzolinoSpoCSpoofingCamera2021, cozzolinoUniversalGANImage2021, gragnanielloAreGANGenerated2021} has demonstrated that this model is able to effectively distinguish real from generated images and that it provides good generalization capabilities.
We initially attempted to use pre-trained fake image detectors, however, we found that the heavy pre-processing performed by Twitter makes it necessary to train our own detector (\cf\ Section~\ref{sec:pretrained_eval}).
In particular, we train on the combination of \datasetLabelRealX and \datasetPseudoLabel for real images, and \datasetLabelFakeX for fake images. The resulting final classifier is denoted by \classifierRxPxFx. Note that we experiment with using other dataset variations to train a classifier in our ablation study in Section~\ref{sec:detector_ablation}. Yet, using processed real, fake, \emph{and} proxy-labeled real images provides the highest performance for processed and zoomed inputs. 

\subsection{Assistance for Manual Labeling}\label{sec:method:assistance}
To estimate error rates of our detection scheme on unlabeled in-the-wild data, it is necessary to manually label these images as real or fake. As generated images have reached a level of quality which makes them almost indistinguishable from real images~\citep{hulzeboschDetectingCNNgeneratedFacial2020, tucciarelliRealnessPeopleWho2020, nightingaleSyntheticFacesHow2021, shenStudyHumanPerception2021, lagoMoreRealReal2022, nightingaleAIsynthesizedFacesAre2022, frankRepresentativeStudyHuman2023}, we use two tools to facilitate this process.

\paragraph{Alignment}
Faces generated by StyleGAN2~\citep{karrasAnalyzingImprovingImage2020} are characterized by being almost perfectly aligned with respect to their facial landmarks, caused by the alignment of the training dataset FFHQ. By superimposing multiple images, this characteristic has been leveraged to visually identify clusters of fake accounts in social networks~\citep{nimmoOperationNavalGazing2020, nimmoIRAAgainUnlucky2020, graphikateamStepMyParler2020, graphikateamFakeClusterBoosts2021, stanfordinternetobservatoryReplyguysGoHunting2020, strickAnalysisProchinaPropaganda2021, goldsteinResearchNoteThis2022}. We automate this manual process by extracting facial landmarks with BlazeFace~\citep{bazarevskyBlazeFace2019} and computing the deviation from a reference. For each landmark $L_1, \dotsc, L_{12}$ (x\nobreakdash-~and y-coordinates of eyes, ears, mouth, and nose), we compute its mean~$\mu_i$ and standard deviation~$\sigma_i$ over 
a reference dataset. In our study, we use the training subset of \datasetLabelFakeX as reference. 
We define an image $x$ as being aligned, if the condition
\begin{equation}
    \vert L_i(x) - \mu_i \vert < k \sigma_i \quad \forall i \in \{1,\dotsc,12\}
\end{equation}
holds, where $L_i(x)$ are the landmarks extracted from the image~$x$ and $k \in \mathbb{Z}$ controls the maximum deviation from the reference.
We set $k=7$. During our evaluation in Section~\ref{sec:eval}, we find that this is the lowest value at which all generated images in the validation set of \datasetLabelFake are aligned.
While a close alignment hints towards a face generated by StyleGAN2, it is ineffective if the image has been cropped or geometrically transformed.

\paragraph{Inversion}
Additionally, we leverage GAN inversion~\citep{xiaGANInversionSurvey2023} as an assistance tool.
For a given input image, this method finds the latent code which reconstructs the original input when passed through the generator.
We use the implementation by~\citet{karrasAnalyzingImprovingImage2020} to invert images using StyleGAN2.
Previous work has shown that generated images can be reconstructed more successfully than real images~\citep{albrightSourceGeneratorAttribution2019,karrasAnalyzingImprovingImage2020,pasquiniIdentifyingSyntheticFaces2023} (we provide a visual example in Appendix~\ref{app:manual_labeling}).
Note that inversion also relies on facial alignment. If an adversary uses a cropped version of a fake face, the inversion result will be distorted. We therefore only use inversion as labeling assistance if the image is aligned.

\section{Evaluation}\label{sec:eval}

In this section, we proceed with an evaluation of our proposed methodology in a controlled setting with labeled data. This allows us to verify the components of our detection pipeline before studying generated faces in the wild on Twitter in Section~\ref{sec:wild} and analyzing the corresponding profiles and tweets in Section~\ref{sec:analysis}.

\paragraph{Dataset Splits}\label{sec:eval:setup}
We randomly split \datasetLabelReal, \datasetLabelFake, and \datasetPseudoLabel into \num{8500} train, \num{500} validation, and \num{1000} test images, respectively. \datasetLabelRealX and \datasetLabelFakeX are split in the same manner. As we use \datasetLabelRealZoomed and \datasetLabelFakeZoomed only for evaluation, they only contain the corresponding \num{1000} test images, respectively.

\begin{table}[tb]
    \centering
    \small
    \caption{Evaluation of our pre-filter~\prefilter. We separately analyze its two conditions, which are the presence of a face and its sufficient size.}
    \label{tab:pre_filtering_eval}
    \begin{tabular}{lrrrr}
        \toprule
        & \multicolumn{2}{c}{With Face} & \multicolumn{2}{c}{Without Face} \\
        \cmidrule(lr){2-3} \cmidrule(lr){4-5}
        Dataset & Face Detected & Size Check & Face Detected & Size Check \\
        \midrule
        \datasetLabelFakeX & \SI{100}{\percent} & \SI{100}{\percent} & 
        --- & --- \\
        \datasetInWild & \SI{92.47}{\percent} & \SI{58.16}{\percent} & \SI{42.53}{\percent} & \SI{30.27}{\percent} \\        
        \bottomrule
    \end{tabular}
\end{table}

\paragraph{Pre-Filter}
We start with the pre-filter~\prefilter that should discard irrelevant images, but keep potentially generated images. An image passes \prefilter if a face (a) is detected and (b) has a sufficient size (see Section~\ref{sec:method:detection}).
In the following, we apply \prefilter to the test set from~\datasetLabelFakeX and to \num{1000} randomly sampled images from~\datasetInWild. For the latter subset, we manually label each image whether it (partly) contains a human face. 
Our experiment here has three goals: we want to verify that all generated images from \datasetLabelFakeX pass \prefilter, confirm that the face detector works reliably on the in-the-wild images from \datasetInWild, and finally get an estimate of the number of kept in-the-wild images passed to the next stage. 

Table~\ref{tab:pre_filtering_eval} shows the results for our evaluation of~\prefilter. All generated images from \datasetLabelFakeX pass \prefilter, fulfilling our first goal.
Among the sampled in-the-wild images from \datasetInWild with a face, the face detector correctly identifies \SI{92.47}{\percent}.
We manually look through the undetected faces. In most cases, the face is either very small, obstructed (\eg by masks or smartphones), or partly outside the frame. The face in these images is not prevalent, so that we consider it acceptable to skip them.
For in-the-wild images without a face, the face detector mistakenly locates a face in \SI{42.53}{\percent} of the cases. We manually inspect the mislabeled images. The vast majority contains faces, but they are drawn, digitally created, or belong to animals or statues. Only very few detections are obviously ``wrong'', such as images with Twitter's former default profile image. Since these images are just passed to the next stage, having some false positives is not critical.
Based on this analysis, we can conclude that the face detector reliably works, fulfilling our second goal.
Finally, we measure how many images from the subset of \datasetInWild additionally pass the size check and therefore \prefilter. In only \SI{58.16}{\percent} of the face images and \SI{30.27}{\percent} of the non-face images, the face is considered large enough, considerably reducing the number of images passed to the next stage. 
Overall, we conclude that our pre-filter allows us to skip irrelevant images efficiently, without mistakenly discarding generated faces.  

\paragraph{Classification}
Next, we verify that our classifier~\classifierRxPxFx is capable of spotting generated images in realistic settings. %
We evaluate the performance of our classifier under three conditions: (a) processed images (\datasetLabelRealX vs.\ \datasetLabelFakeX), (b) zoomed images~(\datasetLabelRealZoomed vs.\ \datasetLabelFakeZoomed), and (c) proxy-labeled real and fake images (\datasetPseudoLabel vs.\ \datasetLabelFakeX). 
We use the test set from each dataset. 

\begin{figure}
    \centering
    \includegraphics{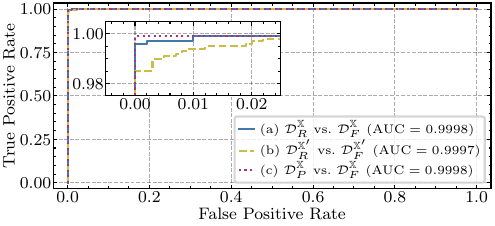}
    \caption{Evaluation of our classifier \classifierRxPxFx. We show the ROC curve under different conditions.}
    \label{fig:roc}
    \Description{ROC curves showing the performance of the proposed classifier. The curves for all three conditions indicate near-perfect performance because they reach the top-left corner of the plot. A close-up shows that the classifiers performs best in condition (c), followed by (a) and (b).}
\end{figure}

Figure~\ref{fig:roc} shows the respective ROC curves. Our classifier has an almost perfect detection rate with an AUC value close to \num{1.0}. Note that the setup on zoomed data is slightly more challenging, because there are no examples of zoomed images in the detector's training data. Still, the error rate remains very small.
Due to the strong class imbalance on real Twitter data, a small error rate is required to avoid an excessive amount of false positives. 

\paragraph{Assistance Tools}
We finally verify our methods that allow us to better label images for the error-rate estimation later.\vspace{4pt}

\noindent\textit{Alignment.}
Using the test set from \datasetLabelFakeX, we confirm that with $k=7$, all fake images are correctly labeled as being aligned.
From the \num{1000} randomly sampled images from \datasetInWild, only 35 are aligned. \vspace{4pt}

\begin{figure}
    \centering
    \includegraphics{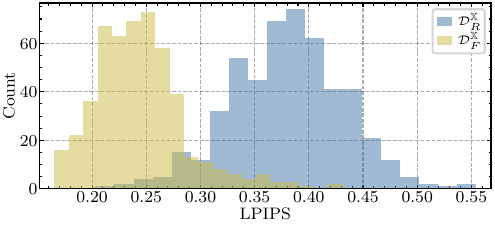}
    \caption{Evaluation of GAN inversion. The lower the LPIPS distance between the original image and its reconstruction, the more similar they are.}
    \label{fig:reconstruction_hist}
    \Description{Two histograms showing the effectiveness of GAN inversion. The LPIPS values of most generated images range from 0.16 to 0.4, while those of real images range from 0.2 to 0.55. Both histograms are mostly separated, there is an overlap around 0.3.}
\end{figure}

\noindent\textit{Inversion.}
We first verify that generated images can be inverted more accurately than real images.
We invert 500 images from \datasetLabelRealX and \datasetLabelFakeX, respectively, and compute the LPIPS~\citep{zhangTheUnreasonableEffectiveness2018} distance between original and reconstructed images.
This distance metric measures the perceptual similarity between two images and has been previously used to estimate the reconstruction quality~\citep{karrasAnalyzingImprovingImage2020}.
The histograms in Figure~\ref{fig:reconstruction_hist} show that the reconstructions from \datasetLabelFakeX are perceptually more similar to the originals compared to the reconstructions from \datasetLabelRealX. A classification based on the LPIPS scores results in an AUC of 0.97.

As second experiment, we check that inversion is helpful for manual labeling.
We divide the \num{1000} images (500 real, 500 fake) into 900 training and 100 test images.
For each image we construct a side-by-side view with the original, its reconstruction obtained by inversion, and the distance measured in LPIPS and MSE (\cf\ Appendix~\ref{app:manual_labeling}).
Using the training set, one annotator practices the manual classification.
We then evaluate the performance based on the held-out test images.
99 out of 100 images are correctly assigned, demonstrating a feasible manual inspection.

We emphasize that images from \datasetLabelRealX are very similar to images from \datasetLabelFakeX.
In contrast, most in-the-wild profile images are visually different, leading to even worse reconstructions (despite being aligned) and thus to comparatively high LPIPS values. Hence, we expect the actual manual labeling process to be easier than in the controlled setting.

\paragraph{Summary}
Our evaluation indicates a valid detection pipeline. The pre-filter allows skipping irrelevant images while the classifier allows detecting generated images. The assistance tools can help with the manual labeling process.

\section{Detecting Generated Images In~the~Wild}\label{sec:wild}
Equipped with a valid detection pipeline, we can now explore the prevalence of generated images on Twitter. To this end, we first need to calibrate the detection to the real-world setup before we can present the final results. 

\paragraph{Manual Labeling}
To begin with, we have to label a subset of~\datasetInWild.
First, this allows us to get a detection threshold, so that we are able to classify individual images as real or generated. Note that in the controlled setup before, we evaluated the overall performance of~\classifierRxPxFx with the AUC metric that takes into account all possible thresholds and thus does not require picking a specific value. 
Second, a separate labeled set is necessary to estimate error rates.

Unfortunately, manual labeling of all samples in \datasetInWild is unfeasible due to the sheer volume of samples within the dataset. Thus, we resort to a random subset of \num{1498938} images, corresponding to \SI{10}{\percent} of all samples.
We then sort these images based on their score (from \classifierRxPxFx) from low (real) to high (fake) and select the top \num{1000} images that pass \prefilter for manual labeling.
We acknowledge that choosing the subset based on the classifier that we are trying to evaluate introduces an unwanted bias: there could be fake images with very low scores that are overlooked.
However, we argue that this approach strikes a balance between practicability and a sound estimation.
Selecting the subset by pure chance would require an enormous amount of manual labeling to gather a sufficient number of fake images.
Moreover, the scores of our subset range from 1.0 to 0.33. From the test set of \datasetLabelFakeX, only 3 out of \num{1000} images get a score below 0.33. We therefore assume that only a very small number of false negatives is potentially overlooked. Our evaluation on the independent dataset \datasetSpotted (at the end of this section) also confirms that our approach provides a reliable estimate of the error rates.

We carefully inspect each image and, if it is aligned, its reconstruction from GAN inversion.
We label an image as real if the framing and pose do not match with that of \datasetLabelFake, if it contains a complex and meaningful background, or if the reconstruction deviates significantly from the original.
In contrast, images are labeled as fake if they contain diffuse backgrounds, asymmetries (eyes, earrings), unnatural clothing, color artifacts, and/or an almost perfect reconstruction.
By doing so, we obtain 185 images labeled as ``Real'', 725 images labeled as ``Fake'', and 90 images labeled as ``Unsure''.
Most images with label ``Unsure'' resemble images from TPDNE, but do not contain clear artifacts or were strongly edited.
We also assigned this label if we suspect that an image was generated using a different kind of generative model. %
We randomly split the 910 images labeled as ``Real'' or ``Fake'' into a validation set (for calibrating the threshold) and a test set (for estimating the error rates) of equal size, maintaining the label ratio in both splits.

\begin{figure}
    \centering
    \includegraphics{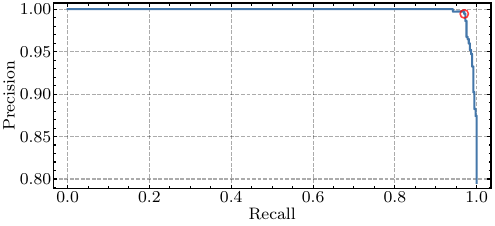}
    \caption{Precision-recall curve of \classifierRxPxFx on the validation set of the manually labeled images. The circle marks the selected threshold, which maximizes the F1-score.}
    \label{fig:precision_recall_val}
    \Description{Precision-recall curve showing the performance of the proposed classifier. It indicates good performance, with precision dropping at a recall value of 0.94.}
\end{figure}

\begin{figure}
    \centering
    \includegraphics{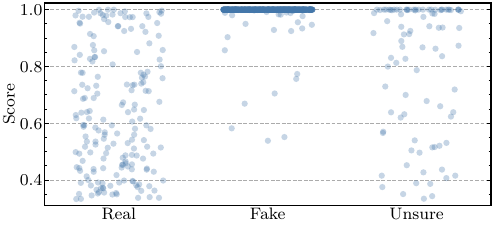}
    \caption{Score distribution of manually labeled images.}
    \Description{Scatter plot showing the classification scores of manually labeled images. The scores of real images are relatively evenly distributed between 0.33 and 1.0. The scores of of most fake images are very close to 1.0, with about 20 outliers. The lowest score is 0.55. Images labeled as ``Unsure'' mostly receive scores of 1.0, but the scores of about 40 images are similarly distributed as real images.}
    \label{fig:manual_scores}
\end{figure}

\paragraph{Choosing a Threshold}
Due to the high imbalance between real and generated images in \datasetInWild, choosing an appropriate threshold is not trivial.
A too high threshold leads to many overlooked fake images (low recall), while a too low threshold leads to many real images classified as fake (low precision).
Figure~\ref{fig:precision_recall_val} depicts the precision-recall curve based on the predictions of \classifierRxPxFx on the validation set of our manually labeled images.
As recall and precision are equally relevant in our setting, we follow the common practice to select the threshold based on the F1-score.
The best F1-score (0.9832) is achieved using a threshold of 0.9899361.
Such a high threshold might appear counterintuitive. 
Figure~\ref{fig:manual_scores} shows that most fake images are confidently classified as fake---with scores very close or equal to 1.
The scores of real images, however, have greater variation.
Thus, choosing a relatively high threshold gives the best performance.
Note that the scores of real images in this subset are not representative for all real images, since we purposely selected images with high scores.

\paragraph{Estimating Error Rates}
Equipped with our selected threshold, we can now estimate the error rates of our detector. We consider two metrics to be most relevant in our setting: the \acf{FNR}, which is the fraction of mislabeled fake images among all fake images, and the \acf{FDR}, which is the fraction of mislabeled real images among \textit{all images classified as fake}. Note that the FDR is not to be confused with the FPR, which is the fraction of mislabeled real images among \textit{all real images}. 

\begin{figure}
    \centering
    \includegraphics{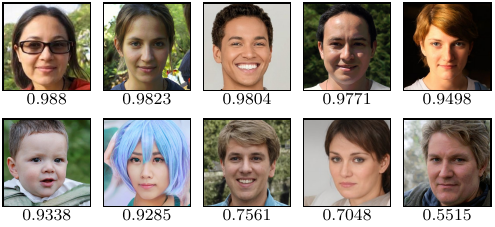}
    \caption{Examples of fake images falsely classified as real, together with their classification score.}
    \label{fig:false_negatives}
    \Description{Ten diverse photos of generated faces that were misclassified.}
\end{figure}

Based on the test set of our manually labeled subset, the FNR is \SI{3.03}{\percent} and the FDR is \SI{1.4}{\percent}.
To understand the errors, we take a closer look on the misclassified images. Figure~\ref{fig:false_negatives} shows the false negatives within the test set together with their scores. Although the majority actually gets a high score and is only classified as real due to the high threshold, three images have a considerably lower score.
We cannot identify a pattern which causes their misclassification.
Neither do we observe any characteristics that would explain the real images classified as fake (false discoveries). As these profile images are real users, we cannot provide visual examples here.

For a second estimate, we leverage the independent dataset~\datasetSpotted of fake profile images that were spotted by users on the web before. We obtain a low FNR of \SI{2.75}{\percent}, that is, \num{39} out of \num{1420} fake profile images are incorrectly labeled as real. This result is similar to our previously estimated FNR of \SI{3.03}{\percent}.
Note that all images in \datasetSpotted pass our prefilter~\prefilter. False negatives therefore only depend on the classifier's score. 

Overall, we can confirm the performance of our detector on two different test sets.
While the error rates are not zero, they are small enough to draw conclusions in our analysis in the next section. 

\paragraph{Prevalence of Fake Profile Images on Twitter}
We are now ready for the final step. We apply our detection scheme on the entire in-the-wild dataset \datasetInWild.
The pre-filter~\prefilter discards \num{8535322} images, reducing the number of images by~\SI{56.94}{\percent}.
Next, using our detector, we classify \num{7723} profile images as fake. This is \SI{0.052}{\percent} of the full dataset.
In the next section, we analyze the profiles behind these images and their tweets in more detail.

\section{Analysis}\label{sec:analysis}
Our goal in this section is to understand the context where the generated profile images are used. To this end, we first perform an analysis of the \emph{accounts} behind these images (Section~\ref{sec:analysis:usermetrics}). Then, we thoroughly analyze the \emph{content of the tweets} that were sent from these accounts (Section~\ref{sec:analysis:contentanalysis}).  
For simplicity, we refer to accounts using generated profile images as ``fake-image accounts'' as opposed to ``real-image accounts'' in the following.

\subsection{User Metrics}\label{sec:analysis:usermetrics}
We begin by analyzing the difference between fake-image and real-image accounts regarding social connections, account activity, as well as account creation and status.

\begin{figure}
    \centering
    \begin{subfigure}[t]{0.49\linewidth}
        \includegraphics{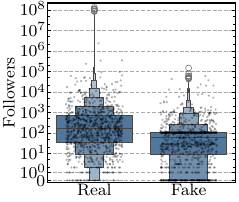}
        \caption{Number of followers.}
        \label{fig:followers}
    \end{subfigure}
    \begin{subfigure}[t]{0.49\linewidth}
        \includegraphics{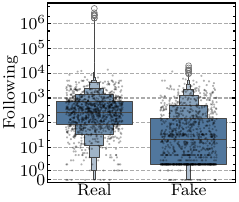}
        \caption{Number of users an account follows.}
        \label{fig:following}
    \end{subfigure}
    \begin{subfigure}[t]{0.49\linewidth}
        \includegraphics{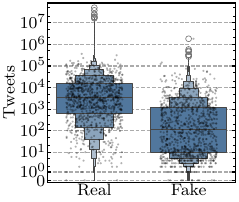}
        \caption{Number of tweets.}
        \label{fig:tweets}
    \end{subfigure}
    \begin{subfigure}[t]{0.49\linewidth}
        \includegraphics{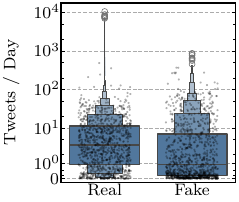}
        \caption{Average number of tweets per day.}
        \label{fig:tweets_per_day}
    \end{subfigure}
    \caption{Distributions of user metrics from real- and fake-image accounts. The points depict \num{1000} randomly selected samples from each class, respectively.}
    \label{fig:metrics}
    \Description{Four boxenplots illustrating the distributions of user metrics from real- and fake-image accounts. Plot (a) shows the number of followers. Real-image accounts have more followers than fake-image accounts. We observe the same behavior for the number of users an account follows in plot (b) and the number of tweets in plot (c). For the average number of tweets per day, shown in plot (d), both distributions are closer together compared to (c).}
\end{figure}

\paragraph{Social Connections}
On Twitter, social interactions are primarily measured in the number of followers an account has and the number of other accounts it follows.
Figures~\ref{fig:followers} and \ref{fig:following} visualize the distribution of these metrics for real- and fake-image accounts at the time of data collection. 
We find that fake-image accounts have fewer followers (mean:~393.35, median:~60) compared to real-image accounts (mean:~\num{5086.38}, median:~165) in our dataset.
This is in accordance with previous analyses of fake social media profiles~\citep{benevenutoDetectingSpammers2010,gurajalaFakeTwitterAccounts2015,singhDetectionFakeProfile2018,cresci-financial}.
\num{1997} (\SI{25.86}{\percent}) of all fake-image accounts have \num{9} or fewer followers and \num{1063} (\SI{13.76}{\percent}) have exactly zero followers.
We notice that \num{1996} fake-image accounts (\SI{25.84}{\percent}) have exactly 106 followers.
Our content analysis in Section~\ref{sec:analysis:contentanalysis} reveals that these accounts belong to a large cluster of fake accounts involved in coordinated inauthentic behavior.

We find that fake-image accounts also follow fewer other accounts (mean: 283.18, median:~21) compared to real-image accounts (mean:~759.83, median:~262).
Interestingly, \num{2175} fake-image accounts (\SI{28.16}{\percent}) follow exactly two other accounts.
In contrast to the number of followers, a relatively small number of fake-image accounts (\num{163}, \SI{2.11}{\percent}) follows exactly zero other accounts.

\paragraph{Activity}
Figure~\ref{fig:tweets} shows that fake-image accounts do participate in Twitter based on the number of tweets. Yet, they are overall less active than real-image accounts.
On average, fake-image accounts posted \num{3158.9} (median:~112) tweets, as opposed to \num{17096.39} (median:~3450) tweets from real-image accounts.
\num{1948} (\SI{25.22}{\percent}) of all fake-image accounts have 10 or fewer tweets.
In addition, Figure~\ref{fig:tweets_per_day} shows the average number of tweets per day, calculated by dividing the total number of tweets by the number of days the account exists.
Based on the median, fake-image accounts are still less active than real-image accounts (0.95 vs.\ 3.68 tweets per day).
However, a large fraction of fake-image accounts posts exceptionally many tweets per day, causing a higher mean (12.89 vs.\ 11.47 tweets per day).
In particular, there are 503 fake-image accounts (\SI{6.51}{\percent}) that submitted more than 50 tweets per day.

\begin{figure}
    \centering
    \begin{subfigure}{\linewidth}
        \includegraphics{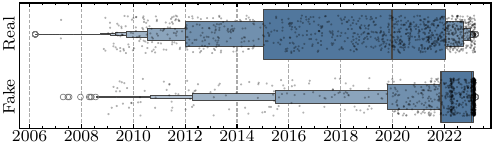}
        \caption{Real- and fake-image accounts.}
        \label{fig:created_at_all}
    \end{subfigure}
    \begin{subfigure}{\linewidth}
        \includegraphics{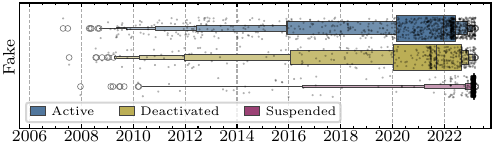}
        \caption{Fake-image accounts split by status.}
        \label{fig:created_at_fake}
    \end{subfigure}
    \begin{subfigure}{\linewidth}
        \includegraphics{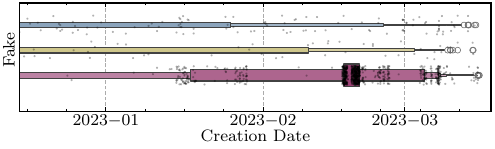}
        \caption{Similar to (b), showing only the last three months.}
        \label{fig:created_at_fake_recent}
    \end{subfigure}
    \caption{Distributions of account creation times from real- and fake-image accounts. In (b) and (c) we differentiate fake-image accounts by their status nine months after data collection. The points depict up to \num{1000} randomly selected samples for each label and status, respectively.}
    \label{fig:created_at}
    \Description{Three boxenplots showing the distribution of creation times for real- and fake-image accounts. Plot (a) compares all real and fake-image accounts, showing that fake-image accounts are considerably younger. Plot (b) shows only fake-image accounts, comparing creation times by account status (acrive, deactivated, suspended). Most suspended accounts were created in 2023, while other two groups are more distributed. Plot (c) shows only the last three months of plot (b), indicating that many suspended accounts were created in short periods of time.}
\end{figure}

\paragraph{Account Creation and Status}
Figure~\ref{fig:created_at_all} compares the times of account creation.
Fake-image accounts are considerably ``younger'', with more than half of them (\SI{52.38}{\percent}) being created in 2023 (note that our data collection happened in March 2023). In contrast, only \SI{6.22}{\percent} of real-image accounts have been created in this period.

\begin{figure}
    \centering
    \includegraphics{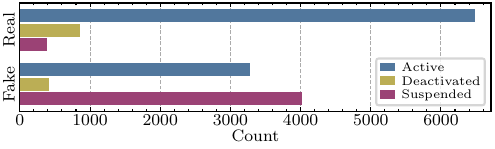}
    \caption{Account status nine months after data collection.}
    \label{fig:status}
    \Description{Horizontal bar plot comparing the account status of real-image and fake-image accounts. More than 6400 real-image accounts are active, while only 800 were deactivated and about 400 suspended. In contrast, 4000 fake-image accounts were suspended and only about 3200 remained active.}
\end{figure}

In addition to the creation date, we also examine the account status after a certain period of time.
We checked the status of all \num{7723} fake-image accounts nine months after data collection by querying the respective profile page.
As a reference, we did the same for an equal number of randomly sampled real-image accounts.
Accounts can be either active, deactivated (by the user), or suspended (by Twitter).
Figure~\ref{fig:status} illustrates that more than half of the fake-image accounts (\SI{52.07}{\percent}) have been suspended.
In contrast, only \SI{5.01}{\percent} of real-image accounts in the reference set have been suspended.
The high number of suspended fake-image accounts suggests that they were violating Twitter's rules.

In Figures~\ref{fig:created_at_fake} and \ref{fig:created_at_fake_recent}, we analyze the account creation of fake-image accounts given their status.
We observe various suspended accounts that were created in bulk just shortly before our data collection, especially in the middle of February.
Note that we do not know when these accounts were suspended, so that we cannot determine the effective lifetime of these accounts.

\paragraph{Takeaways}
Our analysis shows that real and fake-image accounts notably differ. Fake-image accounts have fewer social interactions, both regarding the number of followers and the number of accounts they follow. While these metrics are distributed evenly for real-image accounts, we observe patterns with fake-image accounts. There are large groups with identical values, indicating an orchestrated network of inauthentic users. Moreover, fake-image accounts are not passive, they considerably participate in Twitter based on the number of tweets. Although they are in general less active than real-image accounts, there are several fake-image accounts that post very frequently, hinting towards spamming attacks. 
Finally, fake-image accounts have a more limited lifetime. They are usually created more recently than real-image accounts, and they are also disproportionately often suspended by Twitter. This suggests inauthentic behavior. Moreover, a substantial number was created in bulk just before our data collection period started. This \emph{bulk creation} (or batch creation) is a common pattern for inauthentic behavior, used, for example, to amplify messages or to participate in spamming or trolling activities~\citep{gurajalaFakeTwitterAccounts2015,ferraraTwitterSpamFalse2022}.

\subsection{Content Analysis}\label{sec:analysis:contentanalysis}
To evaluate the purpose of the identified fake-image accounts, we proceed to analyze their tweets (original as well as retweets) posted in 2023. We utilize data collected in the context of a large-scale Twitter stream archiving effort~\citep{fafalios2018tweetskb} based on Twitter's \SI{1}{\percent} sampled stream (the same we used to create \datasetInWild). This allows us to access information about the activity of the profiles before and after the profile collection week (until Twitter restricted access to its API in June 2023). In total, we have access to \num{111165} tweets from the \num{7723} fake-image accounts in our collection. 

\begin{figure}
    \centering
    \includegraphics{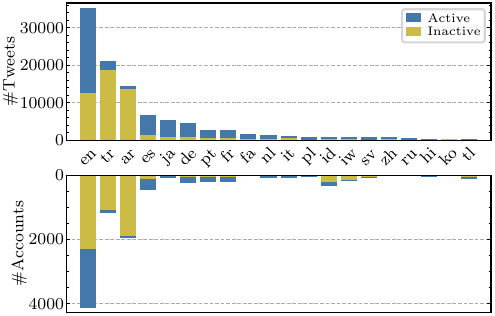}
    \caption{Number of tweets (top) and corresponding accounts (bottom) for the 20 most represented languages. We differentiate accounts (and their tweets) by their status nine months after data collection, combining deactivated and suspended accounts as inactive accounts. Note that a single account can be associated with multiple languages.}
    \label{fig:language_hist}
    \Description{Two vertical bar plots showing the language distribution of fake-image accounts. The upper plot shows the distribution of tweet languages. English is by far the most common language, followed by Turkish and Arabic. The next languages are Spanish, Japanese, German, Portuguese, and French. While less than half of the English tweets come from inactive accounts, the proportion is significantly higher for Turkish and Arabic Tweets. The lower plot shows the corresponding nubmer of accounts per language. More than 4000 accounts tweeted in English, about 1000 in Turkish, and about 2000 in Arabic.}
\end{figure}

We begin our analysis with the \emph{language and account status}.
The upper half of Figure~\ref{fig:language_hist} shows a breakdown of the number of tweets per language. Using the accounts' status nine months after our data collection (\cf\ Section~\ref{sec:analysis:usermetrics}), we can also calculate the fraction of tweets stemming from accounts that have been deactivated or suspended. For the remaining part of this section we refer to this group as inactive accounts. Overall, \SI{49.6}{\percent} of all tweets come from inactive accounts. Interestingly, Turkish and Arabic stand out as languages with significantly higher rates of inactive accounts (\SI{87.95}{\percent} and \SI{95.59}{\percent}, respectively) than other languages. The number of unique accounts that created the tweets in each language are reported in the lower half of Figure~\ref{fig:language_hist}. It shows that Turkish tweets, for instance, stem from a relatively small number of users.

\begin{figure*}[ht]

\includegraphics[width=400px]{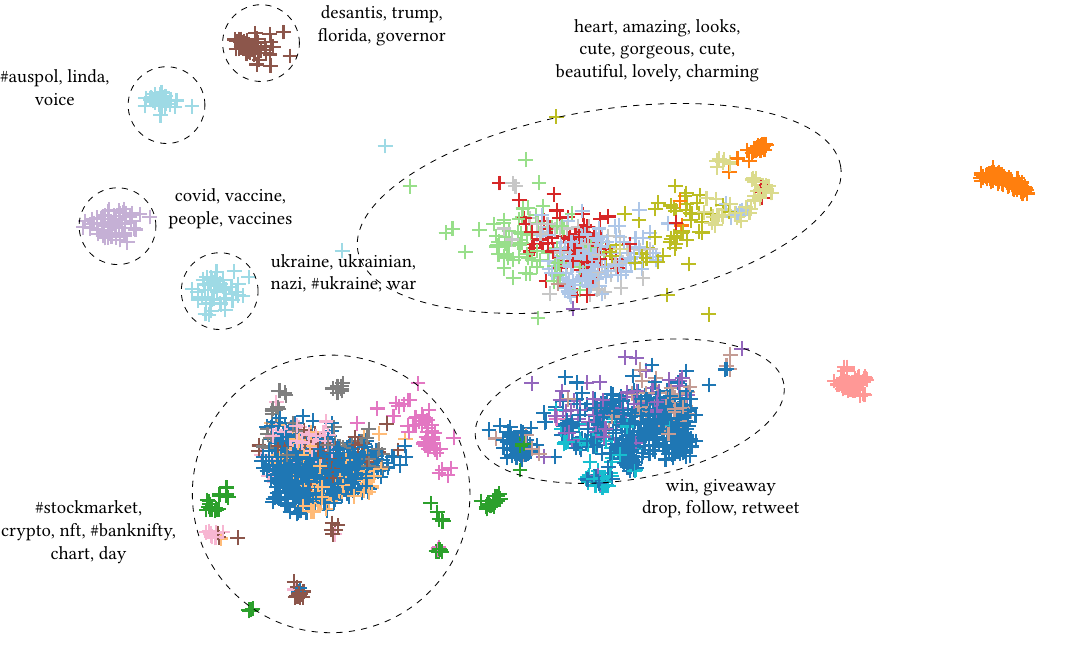}
\caption{UMAP representation of English tweets posted by active users. Distinct (groups of) clusters are annotated by their most representative tokens. 
Different clusters are separated by color.
}
\label{fig:umap}
\Description{UMAP representation of tweet topics. The largest group of clusters is annotated with ``heart, amazing, looks, cute, gorgeous, cute, beautiful, lovely, charming''. Two large clusters are annotated using terms like ``win, giveaway'' and ``\#stockmarket, crypto''. Four smaller clusters are annotated by ``desantis, trump, florida, governor'', ``\#auspol, linda, voice'', ``covide, vaccine, people, vaccines'', and ``ukraine, ukrainian, nazi, \#ukraine, war''.}
\end{figure*}

We proceed with a \emph{textual analysis}. 
To identify structural patterns, we employ state-of-the-art sentence embeddings~\citep{reimers-2019-sentence-bert} to group the tweet texts into semantically related clusters. We utilize the cosine similarity between the sentence embeddings to determine cluster belonging. A new observation (tweet) is assigned to an existing cluster if a certain similarity threshold (in our case 0.6) is reached. Otherwise, a new cluster will be generated. Furthermore, we limit our analysis on clusters that exhibit a minimum cluster size of 50 (i.e., at least 50 tweets should be in one cluster). This approach allows us to identify dominant trends. Note that it does not provide a distribution of topics, because not every tweet is assigned to a cluster. For the purpose of visualization, we use UMAP~\citep{mcinnesUMAPUniformManifold2020} as a dimensionality reduction technique to generate a two-dimensional representation of the clustering outcome (\cf\ Figure~\ref{fig:umap}). For each cluster, we calculate the class-based \ac{TF-IDF} terms to determine representative class tokens. In a subsequent step, we conduct a manual qualitative review of all clusters to identify and describe common themes, which are detailed in the following paragraphs. For translating Turkish and Arabic tweets we use machine translation and double check unclear phrases with native speakers. We describe the general cluster contents and provide representative examples for important topics. We also analyze the metadata of accounts within cluster and report unusual characteristics.

\paragraph{English (inactive accounts)}
The clustering for English content posted by users that are not on the platform anymore reveals a notable pattern: we observe a single, extremely large cluster that encompasses \SI{49.67}{\percent} of all tweets from this group of accounts. Despite the variability of the actual content, these tweets all share a common structure. Each tweet begins by mentioning a specific Twitter user, followed by a short sequence of English terms. Interestingly, these sequences do not form logical sentences, so they are neither semantically nor syntactically correct. Each of these sentences is then followed by a specific Chinese hashtag that can be translated to: ``This is really useful''. Unfortunately, we can only speculate about their purpose. Our hypothesis is that the embedded hyperlinks within the tweets may have directed users to malicious external websites. As the links are no longer functional, we cannot verify this hypothesis. 

The accounts' metadata corroborate the assumption that the \num{1579} accounts within this cluster were part of an organized network. All but three were created between February 16 and February 20, which is consistent with our observations in Figure~\ref{fig:created_at_fake_recent}. Up to 754 accounts were created on a single day. We also find that this cluster contributes to the large number of accounts with identical social connections (\cf\ Figure~\ref{fig:followers} and Figure~\ref{fig:following}). \SI{94.93}{\percent} have exactly 106 followers and \SI{95.31}{\percent} follow exactly two other accounts. The usernames (Twitter handles) appear to be constructed from a list of German-sounding first names and last names (or initials), and optionally one or multiple digits (\eg \textit{@GuntherForstner86}). \SI{67.44}{\percent} of all accounts share the same Arabic display name that can be translated to ``\textit{Noon Namshi Sivvi discount code is strong and effective}'' (Noon and Namshi are e-commerce platforms operating in the Arabic region). These accounts also have their location set to ``KSA'' (Kingdom of Saudi Arabia). Moreover, the accounts contain nonsense descriptions like ``\textit{Personal west service street laugh small.}''. 
We hypothesize that these were automatically generated or translated. Again, we can only speculate 
about the reason, %
especially about the mixed use of English, German, Arabic, and Chinese language. We also notice that \SI{99.18}{\percent} of all accounts in this cluster use profile images that are duplicates within our dataset of fake-image accounts. Appendix~\ref{sec:analysis:duplicates} elaborates our method for identifying duplicate images.

The remaining clusters mainly focus on giveaways, often related to cryptocurrencies, with tweets like
\begin{quote}  
\textit{\$50 (2 winners x \$25)  24 hours  - like, follow)}\\
\textit{i will \#giveaway 100 usdt worth of \$loop as we cele-\hphantom{00}brate our 10k milestone}
\end{quote}
or the promotion of illegal content such as links to broadcasting streams of soccer matches, e.g.,
\begin{quote}  
\textit{live stream arouca vs benfica live [link]}
\end{quote}
Another trend is the distribution of links to websites and Telegram groups containing explicit content, e.g.,
\begin{quote}
\textit{follow for more [link]}
\end{quote}

\paragraph{English (active accounts)}
The clustering of tweets from users who were still active after nine months reveals similarities and differences. Figure~\ref{fig:umap} depicts a visual representation of the top clusters with their representative text tokens. A significant portion of all clusters is again related to various forms of cryptocurrency, stocks, and giveaways, e.g.,
\begin{quote}
\textit{drop your \#tezos \#nft if you need it sold!}\\
\textit{15000\$ in \$eth --- 5 lucky winners!}
\end{quote}

Additionally, we find a significant share of adult content/porn related clusters, actively advertising explicit content, also through dedicated patterns like
\begin{quote}
\textit{beautiful/charming/etc. \@ [profile of porn actress] [link]}
\end{quote}

Compared to inactive users, we observe that active accounts also engage in discussions on contentious or political issues.
These include, for example, the war in Ukraine, election-related discourse, and debates on COVID and vaccinations:
\begin{quote}
\textit{welcome to nazi ukraine \#russia}\\
\textit{desantis racks up wins while trump, potential 2024 \hphantom{00}opponents take swipes at florida governor}\\
\textit{albos crocodile tears: watch this video, that the main-\hphantom{00}stream media refuses to show.}\\
\textit{someone needs to find an antidote for the vaxxx}
\end{quote}

\paragraph{Turkish (inactive accounts)}
For the Turkish accounts, we restrict our analysis to content posted by inactive users, since this is the majority of the dataset. Our findings indicate that almost all of this content is related to pornography or escort services. The primary distinction among the clusters are the cities mentioned within the posts. Most tweets also contain links to other websites, which are no longer functional. Upon examining the metadata of all 932 accounts, we again identify the systematic pattern for usernames that we already observed in the large cluster of English tweets. However, first and last names appear to be of Turkish descent. Moreover, almost all accounts have their location set to a real Turkish city. \SI{46.78}{\percent} of all accounts again use duplicate profile images and \SI{85.52}{\percent} were created within one month. These findings again indicate that at least some systematic approach (automatic or semi-automatic) is used to generate the accounts.

\paragraph{Arabic (inactive accounts)}
For Arabic tweets, we again only consider accounts that have been deactivated or suspended. All clustered tweets appear to be related to literature, with individual clusters being characterized by mentions of certain authors, countries, or topics---all related to the Arabic region. These tweets make up \SI{72.34}{\percent} of all Arabic tweets from inactive accounts. Surprisingly, the tweets share a common structure with those from the large cluster of English tweets: they contain the specific Chinese hashtag, an external link, and an incoherent sentence. Our metadata analysis suggests that the \num{1806} accounts indeed belong to the same cluster, despite the different language. Almost all accounts were created between February 16 and February 20, with 892 being created on a single day. We observe the same anomalies regarding the (German) usernames, locations, descriptions, and social connections. Given the book-related content and the frequently occurring username that promotes a discount code, we hypothesize that the external links might have referred to the respective shopping platforms.

\paragraph{Takeaways}
Our content analysis reveals that English, Turkish, and Arabic are the dominant languages used by the fake-image accounts in our collection. We identify large networks of fake-image accounts that were probably automatically created and that participated in large-scale spamming attacks. 
We observe recurring patterns as part of the automation. Accounts are created in bulk. Tweets, usernames, locations, descriptions, and social connections follow a systematic pattern. Multiple accounts within a network share the same profile image. %
Furthermore, our analysis shows that frequently occurring topics are cryptocurrencies, giveaways, and content related to pornography and escort services. Fake-image accounts also participate in controversial political discussions. These findings align with prior analyses of inauthentic content on Twitter~\citep{ratkiewiczDetectingTrackingPolitical2011,cresciDecadeSocialBot2020,nizzoliChartingLandscapeOnline2020,pfefferJustAnotherDay2023}.

\subsection{Sample Study on Active Accounts}
Finally, we analyze the current behavior of fake-image accounts that are still active at the time of writing (February--March 2024). This gives insights about the use-case of rather long-term fake-accounts. As we cannot use data from Twitter's API any longer, we randomly select \num{1000} active fake-image accounts and visit their Twitter profile manually. Two annotators independently check the most recent tweets and assign a topic to each profile (Cohen's kappa: 0.84). Accounts where both annotators disagree are revisited. We choose topics from five categories, so that we can get a broad understanding of the prevalent application scenarios.

\begin{figure}
    \centering
    \includegraphics{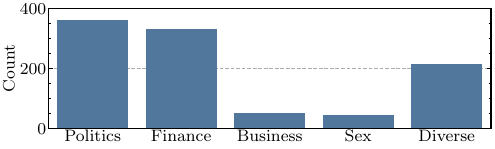}
    \caption{Topic distribution of \num{1000} manually inspected active accounts.}
    \label{fig:manual_topic}
    \Description{Bar plot showing the distribution of topics. The most occurring topics are politics and finance, followed by business and sex. The rightmost bar shows that about 200 accounts tweetes about diverse topics.}
\end{figure}

Figure~\ref{fig:manual_topic} depicts the distribution of topics. The majority of fake-image accounts participates in the political discourse (\SI{36.1}{\percent}) or shares finance-related content (\SI{33.1}{\percent}), mostly related to cryptocurrencies. \SI{5.1}{\percent} of the profiles revolve around other websites or products (``Business''), while \SI{4.3}{\percent} share explicit content or promote escort services (``Sex'').
The remaining accounts (\SI{21.4}{\percent}) cover diverse topics or have an empty timeline.
Taken together, we observe similar topics as before in our cluster analysis. %

\subsection{Summary}
Our systematic analysis revealed \num{7723} Twitter accounts that use AI-generated profile images. By analyzing both their user metrics and the content of their tweets, we identify particular patterns. Some of these patterns, like the high number of suspended accounts, striking similarities within the accounts' properties, or the multitude of similar tweets posted by different users, represent strong evidence that a subset of these accounts is part of organized, inauthentic networks. While many accounts amplify content related to cryptocurrencies or pornography, we also observe accounts that express controversial political opinions.

\section{Ablation Study}\label{sec:ablation_study}
Before finishing our study, %
we shortly confirm the design choices of the classification methodology proposed in Section~\ref{sec:method}. %
In particular, we justify the need to train our own classifier (Section~\ref{sec:pretrained_eval}) and study the impact of training data (Section~\ref{sec:detector_ablation}).

\subsection{Evaluation of Pre-Trained Detectors}\label{sec:pretrained_eval}
Detecting GAN-generated images is a well-researched problem and several pre-trained detectors have been proposed (\cf\ Section~\ref{sec:background:deepfake_detection}). However, we observe that the performance of these detectors suffers from Twitter's image processing, making it necessary to directly train a classifier on processed profile images.

\paragraph{Setup}
We test three existing pre-trained classifiers: 
\classifierWang~\citep{wangCNNgeneratedImagesAre2020} (which is the basis of our classifier),
\classifierGrag~\citep{gragnanielloAreGANGenerated2021}, and
\classifierOjha~\citep{ojhaUniversalFakeImage2023}. 
Appendix~\ref{sec:ablation:pre-trained-detectors} provides more details on these three classifiers. 
We evaluate four conditions (see Table~\ref{tab:detectors_pretrained}). The conditions (a)-(c) correspond to those in Figure~\ref{fig:roc} and all use images processed by Twitter. We additionally test the pre-trained detectors on unprocessed images in condition (d).

\paragraph{Results}
Table~\ref{tab:detectors_pretrained} shows the AUCs of the three classifiers compared to \classifierRxPxFx.
Our trained detector significantly outperforms the pre-trained classifiers under the Twitter conditions (a)-(c). The fact that the latter perform better under the clean condition (d) demonstrates the strong effect of Twitter's processing. It is therefore not possible to use a pre-trained detector for our study of in-the-wild profile images.

\begin{table}
    \centering
    \caption{Evaluation of existing pre-trained detectors. We report the AUCs under different conditions.}
    \label{tab:detectors_pretrained}
    \begin{tabular}{lrrrr}
        \toprule
        Condition                                                   & \classifierWang   & \classifierGrag   & \classifierOjha & \classifierRxPxFx \\
        \midrule
        (a) \datasetLabelRealX vs.\ \datasetLabelFakeX              & 0.7279            & 0.9249            & 0.6405 & 0.9998 \\
        (b) \datasetLabelRealZoomed vs.\ \datasetLabelFakeZoomed    & 0.7243            & 0.9600            & 0.6338 & 0.9997 \\
        (c) \datasetPseudoLabel vs.\ \datasetLabelFakeX             & 0.8713            & 0.9015            & 0.6922 & 0.9998 \\
        (d) \datasetLabelReal vs.\ \datasetLabelFake                & 0.9466            & 1.0000            & 0.8296 & --- \\
        \bottomrule
    \end{tabular}
\end{table}

\subsection{Effect of Training Data}\label{sec:detector_ablation}
Our datasets described in Section~\ref{sec:method:data_collection} allow for different combinations of training data. In the following, we justify the choice of training our detector on real images, proxy-labeled real images, and fake images. 

\paragraph{Setup}
We consider three classifier variants and analyze their performance under the three conditions from Figure~\ref{fig:roc}, respectively. 
The classifier \classifierRF is trained on \datasetLabelReal and \datasetLabelFake and represents the most straightforward option.
The images are not processed by Twitter, but we resize them to $400\times400$ pixels to match the resolution of actual profile images.
The second classifier, \classifierRxFx, is trained on the same images but \textit{with} Twitter's processing.
Finally, \classifierRxPxFx is additionally trained on \datasetPseudoLabel as real images.

\paragraph{Results}
Table~\ref{tab:detectors_auc} shows the AUCs of the three detector variants under the different conditions.
Our finally chosen classifier, \classifierRxPxFx, has the highest performance in all conditions.
The classifier \classifierRF trained on unprocessed images performs worse than both variants trained on processed images, confirming our findings from Section~\ref{sec:pretrained_eval}.
Note that, while an AUC $>0.99$ is still very high, the small difference can cause a significant increase in false positives, given the size of \datasetInWild.
Overall, our results provide two insights. First, the classifier \classifier should be trained on images that are processed similarly to the target images. 
Second, including proxy-labeled real images from the target distribution (\datasetPseudoLabel) improves the detection performance. 
A closer look shows that this causes a better separation of the classifier scores, shifting scores towards either end of the output range.  
This motivates our choice of \classifierRxPxFx for our study.

\begin{table}
    \centering
    \caption{Evaluation of three detector variants trained on different datasets. We report the AUCs under different conditions.}
    \label{tab:detectors_auc}
    \begin{tabular}{lrrr}
        \toprule
        Condition                                               & \classifierRF & \classifierRxFx   & \classifierRxPxFx \\
        \midrule
        (a) \datasetLabelRealX vs.\ \datasetLabelFakeX              & 0.9971        & 0.9998            & 0.9998 \\
        (b) \datasetLabelRealZoomed vs.\ \datasetLabelFakeZoomed    & 0.9953        & 0.9995            & 0.9997 \\
        (c) \datasetPseudoLabel vs.\ \datasetLabelFakeX             & 0.9983        & 0.9994            & 0.9998 \\
        \bottomrule
    \end{tabular}
\end{table}

\section{Discussion and Limitations}\label{sec:discussion}
Our work systematically examines the prevalence of generated images on Twitter.
Despite great effort, our study has limitations that we discuss in the following.

\paragraph{Sampling Bias}
The restricted \SI{1}{\percent} access to Twitter as well as the limited and randomly chosen collection period can introduce a sampling bias~\citep{ArpQuiPen+22} to our study. 
Especially the presence of several large clusters with seemingly orchestrated accounts in our collected dataset has a significant effect on our analysis. These clusters and their concrete topics 
are expected to change over time. 
Nevertheless, the characteristics, such as the bulk creation of accounts, should generally apply. 
The same holds for high-level tendencies, %
such as political amplification or spamming. These are also in line with prior observations on Twitter misuse~\citep{ratkiewiczDetectingTrackingPolitical2011,cresciDecadeSocialBot2020}.
Finally, we note a possible bias due to the restructuring of Twitter/$\mathbb{X}$ after the takeover by Elon Musk. It is possible that with the rise of hate speech and bots~\citep{hickeyAuditingElonMusk2023}, the prevalence of generated profile images has also increased. Unfortunately, the current API limits impede a replication of our analysis.

\paragraph{Selection Bias in Analysis}
A full analysis of all tweets is beyond the scope of our work. Thus, our cluster analysis is not exhaustive and only focuses on the prevalent trends. Still, this allows us to identify the primary contexts in which generated images are used on Twitter, so that we can draw general conclusions on topics.

\paragraph{Focus on Images Generated from TPDNE}
We focus on facial images from TPDNE that are generated by StyleGAN2~\citep{karrasAnalyzingImprovingImage2020}. Although we are unable to provide statements regarding the prevalence of other types of generated images on Twitter, we expect to cover the most prevalent type.  
TPDNE has made it considerably easier to access generated images compared to other generative models. Several reports confirm that GAN-generated faces are in fact used by fake social media accounts~\citep{nimmoOperationFFSFakeFace2019,nimmoOperationNavalGazing2020, nimmoIRAAgainUnlucky2020, graphikateamStepMyParler2020, stanfordinternetobservatoryReplyguysGoHunting2020,nimmoSpamouflageGoesAmerica,stanfordinternetobservatoryAnalysisTwitterTakedowns2020,strickAnalysisProchinaPropaganda2021,graphikateamFakeClusterBoosts2021,williamsPortraitModeGAN2022}. Moreover, most alternative models need to be deployed locally. This requires technical knowledge and possibly specialized hardware. Although text-to-image models like Stable Diffusion or Midjourney can be accessed through a browser, generating images at scale may require significant time and additional costs. Achieving good images can require multiple attempts and services like Midjourney require payment.
Finally, we note that detecting \textit{all} kinds of generated images, especially in a real-world setting where images are heavily processed, is still an open challenge~\citep{gragnanielloAreGANGenerated2021,corviDetectionSyntheticImages2023}. Therefore, we focus on one setting where we aim at developing a highly reliable detector.

\paragraph{Likelihood of Overlooked Fake Profiles}
As discussed in Section~\ref{sec:wild}, classifying in-the-wild data always requires trading off the number of overlooked fakes against the number of falsely detected real images. While we make our best efforts to evaluate the performance of our detection pipeline under realistic conditions, we cannot exclude that the actual FNR is higher than our estimate. Fake profile images with an unusual processing could potentially bypass our detector. Furthermore, our tool-assisted manual labeling process is not guaranteed to be error-free. However, as the FNR on the independent dataset \datasetSpotted closely matches our estimate, the likelihood of overlooked generated images should be low. 

\paragraph{Applicability for Fake Account Detection}
While our proposed detection pipeline can reliably identify accounts using generated profile images, this is naturally not sufficient for mitigating fake social media accounts in general. Not only does a generated profile image not necessarily imply that an account is fake, inauthentic accounts can simply use other image sources. State-of-the-art fake account detection methods usually rely only on account metadata and social interactions~\citep{yangScalableGeneralizableSocial2020}. However, we believe that profile images can be an additional factor to decide whether an account is authentic or not. While stock photos or images taken from the Internet can be easily identified using reverse image search, generated images have the advantage (for the adversary) of being unique. Our work can contribute to closing this loophole.

\section{Related Work}\label{sec:related_work}
Studying generated faces on social media touches different research areas. In the following, we examine related methods and concepts. 

\paragraph{Detecting Generated Images on Social Media}
Despite the plethora of proposed fake image detection methods (\cf\ Section~\ref{sec:background}), there exists only little work on the detection in real-world settings.
\citet{boatoTrueFaceDatasetDetection2022} create a synthetic dataset of processed images by sharing real and generated images on different social media platforms.
They find that a classifier trained on ``original'' images is not able to effectively detect shared images, unless it is fine-tuned. This confirms our results in Section~\ref{sec:ablation_study}.
In a related work~\citep{marconDetectionManipulatedFace2021}, the same methodology is applied to deepfake videos, yielding similar findings.
\citet{sabelDetectingGeneratedMedia2021} present an approach %
to detect generated text and profile images on Twitter.
They collect tweets related to controversial topics (e.g., COVID-19) and separately classify the tweet's text and the corresponding profile image.
Their method can detect generated media but is highly sensitive to selected thresholds.
High precision thresholds cause a significant decrease in true positives, resulting in many overlooked generated images.

Closest to our work is the concurrent study by \citet{yangCharacteristicsPrevalenceFake2024}. 
While they share the goal of estimating the prevalence of generated profile images on Twitter, their dataset, methodology, and analysis is more limited. The number of analyzed in-the-wild profile images is considerably smaller (\num{254275} vs.\ our \num{14989385} samples). 
Moreover, our detection approach consists of multiple stages to obtain a reliable and robust detection. In contrast, the detection approach by \citet{yangCharacteristicsPrevalenceFake2024} is solely based on StyleGAN2's facial alignment, which is not sufficient to effectively identify generated faces. First, some real faces inevitably have the same facial alignment. This explains that their evaluation shows a high FDR of \SI{85.86}{\percent}. Thus, each detected image needs to be manually checked, which does not scale with a larger dataset. Second, relying exclusively on facial alignment is also vulnerable to simple geometric transformations like scaling or cropping. After applying the approach on \datasetInWild, we find that 440 generated images in \datasetInWild cannot be detected due to being misaligned. We provide a detailed description of their method and the results of our comparison in Appendix~\ref{sec:ablation:ganed}.
Finally, in contrast to \citet{yangCharacteristicsPrevalenceFake2024}, we contextualize the use of generated profile images by examining the corresponding accounts and tweets.

\citet{yangCharacteristicsPrevalenceFake2024} estimate a lower bound of \SI{0.021}{\percent} to \SI{0.044}{\percent} active Twitter accounts that use generated profile images. Our estimated rate with \SI{0.052}{\percent} is slightly higher, which we attribute to our higher detection performance and the fact that we discard accounts with Twitter's default profile image.

\paragraph{Human Perception of Generated Social Media Profiles}
Since it is unlikely that artificially generated social media profiles can be prevented completely, studying their effect on humans and our society is crucial.
\citet{minkDeepPhishUnderstandingUser2022} conduct a user study to measure users' trust towards such profiles in a social engineering context.
They find that users are likely to accept a connection request from a LinkedIn profile using generated faces or texts.
Even participants that were explicitly informed about the presence of fake accounts had an acceptance rate of \SI{43}{\percent}.
A similar work~\citep{rossiAreDeepLearninggenerated2023} in which participants were asked to label profiles as real or fake in a Twitter-like environment, shows that human performance is almost equivalent to random guessing (\SI{48.9}{\percent}).
These findings emphasize the need for reliable detection methods of generated contents in social networks.

\paragraph{Social Media Studies} %
Complimentary to our work, a large body of interdisciplinary research has focused on the misuse of social media~\citep{cresciDecadeSocialBot2020, ferrara_challenges, yardi2010detecting}. 
For example, the 2016 US elections were marked by accusations of opinion manipulation through automated accounts on social media, particularly on Twitter, so that researchers investigated these inauthentic and coordinated campaigns~\citep{bessi2016social,badam}. In recent years, research has increasingly focused on the harms caused to online communities and the potential to manipulate public sentiment. Studies have extensively explored the roles of disinformation spread, online conspiracy proliferation, and political interference~\citep{wang2023,shao2018spread,luceri2019evolution}. Another research direction is the identification of inauthentic behavior in context of financial campaigns~\citep{cresci-financial, tardelli}. 
Recently, these research efforts are facing new challenges given the increasing use of AI-generated content by social bots~\citep{ferrara_challenges, Yang_Menczer_2024}.

\section{Conclusion}\label{sec:conclusion}
Generative AI provides unprecedented capabilities to create deceptively realistic content, be it images, videos, text, or music. Despite the considerable applications for the good, these methods also raise significant concerns about their harmful effects. On social media, generated images can be misused to create seemingly real accounts that spread, for example, political misinformation or spam.    
While the detection of generated content has been explored extensively in controlled laboratory settings, there has been limited systematic research on the prevalence on social media. %
In this paper, we provide the first systematic large-scale study of generated profile images on Twitter. 
To build a reliable detection method, we carefully build a pipeline step by step where we consider different dataset types, pre-filtering, classification, and labeling-assistance methods. 

In our dataset of \num{14989385} profile images from Twitter, we classify \num{7723} profile images as generated. This is \SI{0.052}{\percent} of the dataset, showing that generated profile images are notably present on Twitter.   
Our analysis of the corresponding accounts and their tweets leads to various insights. Fake-image accounts and real-image accounts differ regarding social connections, account activity, account creation time, and account status. 
For example, many fake-image accounts are created in batches and have identical metadata, indicating that they are part of an organized network.
The tweet analysis shows that frequently occurring topics are cryptocurrencies, giveaways, content related to pornography and escort services, as well as controversial political discussions. 

In summary, our work introduces a detection method for studying generated content on social media. Our analysis underlines that generated images are used as profile images for a wide range of applications. 
Addressing this threat will require several steps. First, platforms can implement detection algorithms to flag generated content, as Meta has announced lately~\citep{cleggLabelingAIgeneratedImages2024}. 
Second, watermarking methods (\eg\ \citep{fernandezStableSignatureRooting2023}) that integrate a detectable watermarking directly into the generation process can facilitate the detection. Finally, raising more awareness about the existence and impact of generated content will be necessary.

\section*{Ethics Statement and Data Availability}
Working with real-world data from social media carries ethical and privacy-related risks. We take different measures to reduce these risks. In our study, statistics of real accounts are reported in aggregated form. We show personal information, such as profile images and tweet texts, only for accounts using generated images. However, we acknowledge that we cannot completely avoid the risk of falsely labeling a real image as generated. 

To foster the development and evaluation of real-world generated image detectors, we plan to share our labeled image datasets.
Moreover, to comply with Twitter's/$\mathbb{X}$'s terms of service (ToS), we will release the IDs of users and tweets from our in-the-wild dataset. Due to the recent changes to Twitter's API, we are aware that accessing the full dataset based on the IDs is challenging. We therefore invite researchers to contact us for discussing further uses of the dataset and potential collaborations. %

\begin{acks} %
We thank our reviewers for their valuable comments and suggestions.
This work was funded by the Deutsche Forschungsgemeinschaft (DFG, German Research Foundation) under Germany's Excellence Strategy - EXC 2092 CASA - 390781972.
Moreover, this work was supported by the German Federal Ministry of Education and Research (BMBF) under the grant UbiTrans (16KIS1900), the Leibniz Association Competition (P101/2020), as well as by the IFI program of the German Academic Exchange Service (DAAD) funded by the Federal Ministry of Education and Research (BMBF). 
\end{acks}

\bibliographystyle{ACM-Reference-Format}
\bibliography{main}

\appendix
\section{Methodology Details}\label{app:implementation}
Here, we provide implementation details of our methodology. 

\subsection{Data Collection}\label{app:implementation:data}
\paragraph{In-the-Wild Dataset \datasetInWild}\label{app:metadata}
We access the Twitter API using the tweepy~\citep{tweepy} Python package and download the profile image of each tweet's author from the respective \texttt{profile\_image\_url}.
Table~\ref{tab:metadata} lists all metadata fields we obtain from the API. The second column denotes how many accounts in \datasetInWild have a value in the respective field.

\renewcommand{\arraystretch}{1.0}
\begin{table*}
    \centering
    \caption{Overview of metadata for accounts in our dataset~\datasetInWild.}
    \label{tab:metadata}
    \begin{tabular}{lrl} \toprule
         Field & Count & Description \\ \midrule
         \texttt{id} & \num{14989385} & Unique user identifier. \\
         \texttt{username} & \num{14989385} & Username (handle). \\
         \texttt{name} & \num{14989385} & Name shown in profile (display name). \\
         \texttt{created\_at} & \num{14989385} & Account creation time. \\
         \texttt{location} & \num{7940863} & User-specified location. \\
         \texttt{description} & \num{14989385} & Profile bio. \\
         \texttt{url} & \num{3654749} & User-specified URL. \\
         \texttt{profile\_image\_url} & \num{14989385} & URL to user's profile image. \\
         \texttt{public\_metrics.followers\_count} & \num{14989385} & Number of followers. \\
         \texttt{public\_metrics.following\_count} & \num{14989385} & Number of accounts user is following. \\
         \texttt{public\_metrics.tweet\_count} & \num{14989385} & Number of tweets. \\
         \texttt{public\_metrics.listed\_count} & \num{14989385} & Number of lists containing user. \\
         \texttt{protected} & \num{14989385} & Whether account is private. \\
         \texttt{verified} & \num{14989385} & Whether account is verified. \\
         \texttt{withheld.country\_codes} & \num{4192} & Countries where user is not available. \\
         \texttt{pinned\_tweet\_id} & \num{6866224} & Identifier of user's pinned tweet. \\
         \texttt{entities.url.urls} & \num{3654749} & Details about profile website. \\
         \texttt{entities.description.mentions} & \num{1611345} & Details about user mentions in description. \\
         \texttt{entities.description.urls} & \num{780440} & Details about URLs in description. \\
         \texttt{entities.description.hashtags} & \num{1340520} & Details about hashtags in description. \\
         \texttt{entities.description.cashtags} & \num{35746} & Details about cashtags in description. \\
         \bottomrule
    \end{tabular}
\end{table*}

\paragraph{Labeled Datasets \datasetLabelReal/\datasetLabelFake and Variations}
We collect \num{10000} images from TPDNE by repeatedly querying the website, mimicking a user creating a fake profile.
We analogously take the first \num{10000} real images from the FFHQ dataset.
To avoid an unwanted bias based on image processing, we convert the PNG files from FFHQ to JPEG using the same parameters as TPDNE.
Then, to obtain processed images as they would appear on Twitter (\datasetLabelRealX and \datasetLabelFakeX), we upload each image as a profile image and download it.  
We observed a difference in the image processing between API-based and browser-based uploads. Images uploaded with the API kept their resolution, while images uploaded in the browser were resized to $400\times400$ pixels.
As the majority of in-the-wild images has the resized resolution, we select the browser-based approach and automate the upload using the web automation framework Selenium~\citep{selenium}.
To obtain the zoomed-in versions (\datasetLabelRealZoomed and \datasetLabelFakeZoomed), the automated upload procedure is extended by first zooming into each image by a random amount and then moving the image by a random x- and y-offset. We ensure that the image still looks like a plausible profile image at the maximum zoom rate.

\subsection{Pre-Filter}\label{app:implementation:prefilter}
BlazeFace~\citep{bazarevskyBlazeFace2019} predicts a bounding box as well as the x- and y-positions of six facial landmarks (eyes, ears, mouth, and nose) in normalized coordinates between 0 and 1. If an image contains multiple faces, we select the one with the largest bounding box.

\subsection{Classifier}\label{app:implementation:classifier} %
\label{sec:appendix-eval-setup}
Our architecture and training procedure is adapted from \citet{wangCNNgeneratedImagesAre2020}.
We follow the common practice of initializing a ResNet-50~\citep{heDeepResidualLearning2016} with weights from an image classifier trained on ImageNet~\citep{russakovskyImageNetLargeScale2015} and replace the final layer to reflect the binary classification setting.
During training, we use a batch size of \num{32} and optimize the model using  Adam~\citep{kingmaAdamMethodStochastic2015} and binary cross-entropy loss.
In the case of \classifierRxPxFx we ensure balanced sampling of real/proxy-labeled real and fake samples.
The learning rate is reduced by a factor of 10 if the validation loss does not decrease by \num{0.001} during 5 epochs. We perform early stopping once the learning rate becomes smaller than $10^{-6}$. For training \classifierRF, the images in \datasetLabelReal and \datasetLabelFake are resized to $400\times400$ using bilinear interpolation to match the profile image dimensions of Twitter. The training data is augmented using three kinds of perturbations, each applied with probability $p=0.1$: Gaussian blurring with a kernel size of \num{9} and $\sigma$ uniformly sampled from $[0.5, 5.0]$, JPEG compression with quality uniformly sampled from $[30, 100]$, and resizing, with scale and aspect ratio uniformly sampled from $[0.25, 0.75]$ and $[0.8, 1.25]$, respectively. During training, we randomly extract crops of size $224\times224$, while we take the center crop of the same size during validation and testing.

\section{Inversion Examples}\label{app:manual_labeling}
\begin{figure}
    \centering
    \begin{subfigure}{0.8\linewidth}
        \includegraphics[width=\linewidth]{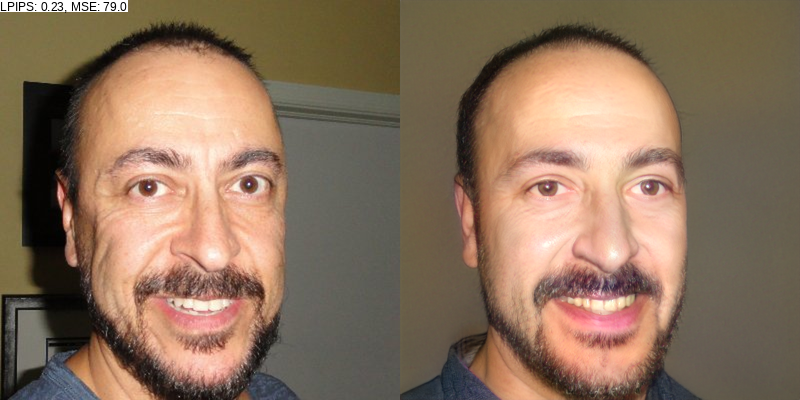}
        \caption{Real image from \datasetLabelReal.}
    \end{subfigure}
    \begin{subfigure}{0.8\linewidth}
        \includegraphics[width=\linewidth]{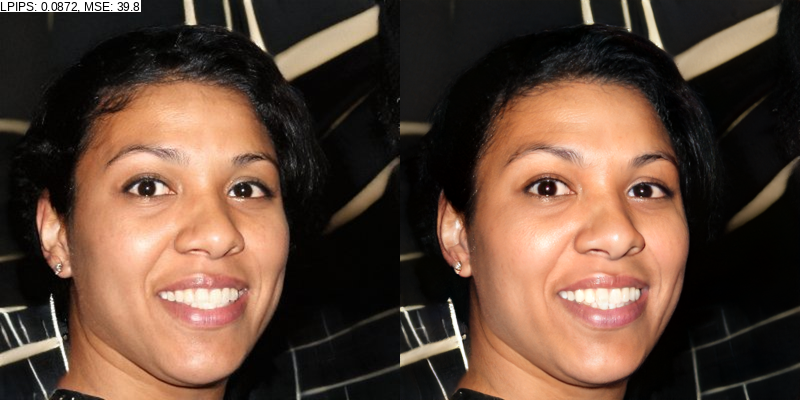}
        \caption{Generated image from \datasetLabelFake.}
    \end{subfigure}
    \caption{Examples of original images (left) and their reconstructions (right).}
    \label{fig:inspection_example}
    \Description{Two images depicting original and reconstruction of a real and generated face, respectively.}
\end{figure}

Figure~\ref{fig:inspection_example} depicts example images to demonstrate the assisted manual labeling. The left image is the original while the right image is its reconstruction obtained by GAN inversion. For the real image from \datasetLabelReal, we observe that the background is inaccurate and the face is slightly blurred. In contrast, the generated image from \datasetLabelFake can be reconstructed very accurately, including the background.

\section{Pre-Trained Detectors}\label{sec:ablation:pre-trained-detectors}
Here we provide details on the three existing pre-trained classifiers we evaluate in Section~\ref{sec:pretrained_eval}.
\classifierWang~\citep{wangCNNgeneratedImagesAre2020} is the model on that our detector is based on.
However, it is trained on a diverse set of images generated by ProGAN~\citep{karrasProgressiveGrowingGANs2018} and corresponding real images from LSUN~\citep{yuLSUNConstructionLargescale2016}.
We select the version \textit{Blur+JPEG (0.1)} since the authors report a good performance on images generated by StyleGAN2~\citep{karrasAnalyzingImprovingImage2020}.
\classifierGrag~\citep{gragnanielloAreGANGenerated2021} is an improved version of \classifierWang that avoids downsampling in the first layer of the ResNet-50~\citep{heDeepResidualLearning2016} backbone to preserve high-frequency artifacts (at the cost of a larger model).
Besides training on ProGAN~\citep{karrasProgressiveGrowingGANs2018} images, the authors provide a detector trained on StyleGAN2~\citep{karrasAnalyzingImprovingImage2020} images, which we select since it should yield the best results on our dataset.
Finally, \classifierOjha follows a different approach and leverages the feature space of a pre-trained vision transformer (CLIP-ViT~\citep{dosovitskiyImageWorth16x162020, radfordLearningTransferableVisual2021}.
It uses a single linear layer on top (trained on ProGAN~\citep{karrasProgressiveGrowingGANs2018} images) to predict whether an image is real or fake.

\section{Duplicate Image Detection}\label{sec:analysis:duplicates}
Despite the trivial access to generated faces using TPDNE, creators of fake account clusters might use the same face for multiple accounts. To identify such duplicates, we need an approach that is robust to subtle differences caused by varying image processing. 
We adapt the technique used by previous works~\citep{garimellaImagesMisinformationPolitical2020,zannettouOriginsMemesMeans2018,wangUnderstandingUseImages2023} and cluster images based on their perceptual hashes (pHashes).
Perceptual image hashing~\citep{faridOverviewPerceptualHashing2021} aims to extract a meaningful representation of an image that does not depend on individual pixel values, but on the perceived content.
The algorithm we use~\citep{buchnerImageHash} achieves this by deriving 64~bits from the DCT coefficients belonging to the lower frequencies of an image.
To obtain groups of duplicate images, we apply the DBSCAN~\citep{esterDensitybasedAlgorithmDiscovering1996} clustering algorithm to our calculated pHashes.
We use the implementation from scikit-learn~\citep{scikit-learn} and set the minimum number of elements to 2.
We empirically find that we obtain meaningful clusters by setting the maximum allowed Hamming distance between two pHashes to 3.

\begin{figure}
    \centering
    \includegraphics{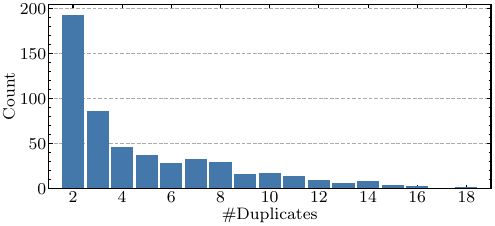}
    \caption{Size distribution of duplicate image clusters.}
    \label{fig:cluster_hist}
    \Description{Histogram showing the distribution of duplicate image cluster sizes. Clusters with size 2 are by far most common, with about 190 clusters. There are about 85 cluster of size 3. The distribution steadily decreases towards size 18.}
\end{figure}

In total, we identify 540 groups of duplicated images with an average size of 4.88 images.
The distribution of the sizes is given in Figure~\ref{fig:cluster_hist}.
About half of all groups consist of only two or three duplicated images, while the most frequently used faces appeared in 18 profiles.

\section{Evaluation of Alignment-Based Detection}\label{sec:ablation:ganed}
In the concurrent work by \citet{yangCharacteristicsPrevalenceFake2024}, the authors identify GAN-generated faces on Twitter using a method that is related to our concept of alignment (\cf\ Section~\ref{sec:method:assistance}).
They define the \textit{GANEyeDistance} $\mathcal{G}$ as the normalized Euclidean distance between the actual and expected location of each eye.
They propose to consider an image to be potentially GAN-generated if $\mathcal{G} < 0.02$.
To reach a final decision, they propose to manually classify images based on visual artifacts.
While this approach is easy to implement and computationally efficient, we find that is suboptimal regarding (a) the number of false positives (causing a large manual workload) and (b) the number of false negatives (overlooking generated faces that are not aligned).

We test $\mathcal{G}$ with the suggested threshold on \num{150000} randomly chosen images from \datasetInWild (about \SI{1}{\percent}, which yields 730 candidate profiles.
For all samples in \datasetInWild, the estimated number of candidates therefore is \num{73000}.
Classifying these images would require an excessive amount of manual effort.

\begin{figure}
    \centering
    \includegraphics{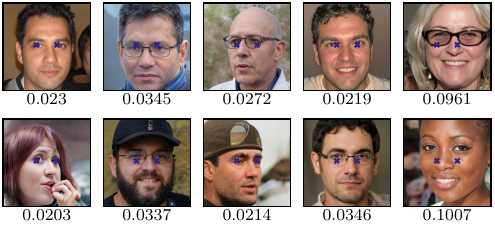}
    \caption{Examples of fake images that evade alignment-based detection. Below each image we provide its \textit{GANEyeDistance} $\mathcal{G}$. The reference eye position is highlighted. Note that we only display images which we confidently consider to be fake to avoid disclosing real profile images.}
    \label{fig:ganed_evasion}
    \Description{Ten photos of generated faces. Two crosses mark the reference eye position. For all images, the true eye position deviates from the reference.}
\end{figure}

On the other hand, we find 440 images in \datasetInWild that are detected as fake by \classifierRxPxFx but are overlooked when classifying based on~$\mathcal{G}$.
Naturally, it can be assumed that in this subset our classifier has a higher number of false positives, since most generated images are in fact aligned.
Still, after manual inspection, we rate 303 of these images to be definitely or very likely generated.
Note that manual labeling is more challenging on these images since we cannot resort to GAN inversion.
Figure~\ref{fig:ganed_evasion} depicts some examples together with their value of $\mathcal{G}$.
One can see that zooming in by a small amount is sufficient to cause a misalignment.
We consider it probable that malicious accounts do this on purpose to appear more credible and avoid detection based on facial landmarks.

\end{document}